# Determining the j-Specific Unoccupied 5f Electronic Structure in Actinides

J. G. Tobin
Department of Chemistry and Physics,
University of Wisconsin-Oshkosh, Oshkosh, WI, USA 54901, USA
Email: tobinj@uwosh.edu

**Abstract**

An empirical comparison of high resolution $M_{4,5}$ X-ray Absorption Spectroscopy (XAS), based upon High Energy Resolution Fluorescence Detection (HERFD), with Bremsstrahlung Isochromat Spectroscopy (BIS), the high energy variant of Inverse Photoelectron Spectroscopy (IPES), of various actinide materials is presented here. It will be shown that XAS-HERFD is essentially the j-specific version of BIS, providing a direct measure of the separate $5f_{5/2}$ and $5f_{7/2}$ Unoccupied Density of States (UDOS).

# Determining the j-Specific Unoccupied 5f Electronic Structure in Actinides

## Introduction

The actinides are problematic, both technically and scientifically. A significant fraction of modern society's electrical power is generated by nuclear fission reactors, particularly those using uranium dioxide as a fuel, e.g., with the United States and Russia at 18% and France, Slovakia, Hungary and Ukraine near or above 50% in 2024. [1] However, the disposal of nuclear waste products remains troublesome. [2,3] Scientifically, the 5f electronic structure resists the ongoing efforts to unravel its enigmas. [4] There are several reasons for this.

First, the proper implementation of fully relativistic calculations has only occurred relatively recently. [5-13] Historically, there have been many treatments of various elemental forms of the actinides and their compounds and alloys. [14] Unfortunately, without a modifying influence, the tendency in simple DFT (Density Field Theory) and related methods is to make the 5f bands far too dispersive and delocalized. [15] One result of this was to set up the situation where artificially magnetic calculations were the most accurate. [16] A turning point was the implementation of methods of Kutepov [5] and Soderlind. [7]

Second, the 5f Photoelectron Spectroscopy (PES) measurements of actinides have had limited success. A contributing factor here is the difficulty in preparing pure samples and keeping the surfaces clean, with surface lifetimes on the scale of hours, worsened by the issues of the radioactivity of the samples. Moreover, in the early actinides, only a few of the 5f states are actually occupied and the spectra tend to be all much the same, with a peak at the Fermi

Energy or Valence Band Maximum and intensity dropping off sharply as the binding energy increases. [17] Even the angle (momentum) resolving variants of Photoelectron Spectroscopy have related constraints, with only small 5f dispersion. [18]

Third, 5f Inverse Photoelectron Spectroscopy (IPES)and Bremsstrahlung Isochromat Spectroscopy (BIS) measurements have similar problems, worsened by the factor of 10,000 reduction in intensity relative to PES. [19] Nevertheless, the IPES/BIS measurements have greater potential, with many more unoccupied states in the early actinides. However, previous analyses of the BIS of 5f materials were hampered by the history of success of the analyses of 4f materials. The success of Russell-Saunders-based interpretations of 4f systems were so overwhelmingly effective, they skewed the interpretation of the actinide 5f electronic structure. [20,21]

Fourth, until recently, all X-ray Absorption Spectroscopy (XAS) of 5f Unoccupied Density of States (UDOS) were hampered by the large lifetime broadenings of the 5d-5f, 4d-5f and 3d-5f transitions. [22] Despite these large lifetime broadenings, it was shown with the XAS and Branching Ratios (BR) of Pu that Russell-Saunders interpretations fail for Pu and that the Intermediate Coupling Model, skewed toward a jj-coupling picture, is correct. [23-26] Nevertheless, the large lifetime broadenings of the XAS rendered moot any attempt to directly measure the detailed structure of the 5f UDOS with XAS, until very recently. [27]

# Determining the j-Specific Unoccupied 5f Electronic Structure in Actinides

With the development of the High Energy Resolution Fluorescence Detection (HERFD) in XAS by Kvashnina et al [28] and Vitova et al [29], the situation for XAS has changed completely. In essence, HERFD reduces the lifetime broadening by utilizing an alternative X-ray Emission detection channel. To be exact, this means that the experiment is now an X-ray Scattering investigation [30] with the possibility that other non-absorption effects can creep into the results [31]. Despite this, there is now a significant accumulation of results to support the contention that HERFD is indeed an XAS experiment. [30, 32-36]

The improved resolution of the HERFD experiment, with a band pass of less than 1 eV, opens new doors to the direct interrogation of the 5f UDOS. The previous success of the combination of the 4d-5f XAS BR experiments at about 800e eV and the Intermediate Coupling Model (ICM) is proof of the accuracy of the electric dipole rules in these transitions. [24-26] In HERFD, the Tender X-ray (~ 3000 eV) 3d-5f transitions are used, and it is possible that the 3d-5f electric dipole transition selection rules and moments may be compromised by higher order effects, e.g. electric quadrupole transitions. However, there are now more data from the Tender X-ray regime, which indicate that the electric dipole transition selections rules and moments still hold, e.g. the observation of angular momentum coupling effects in the X-ray Emission of Uranium Tetrafluoride [37-39] and the recent analyses of the actinide dioxides. [32] Thus, it is possible to view the uranium $M_4$ and $M_5$ HERFD experiments as direct measurements of the $5f_{5/2}$ and $5f_{7/2}$ UDOS. [32-36, 40]

# Determining the j-Specific Unoccupied 5f Electronic Structure in Actinides

Historically, an accepted procedure to determine the Unoccupied Density of States (UDOS) was and is Bremsstrahlung Isochromat Spectroscopy (BIS), the high energy variant of Inverse Photoelectron Spectroscopy (IPES). [14, 41-56] (Figure 1) Just as X-ray Photoelectron Spectroscopy can provide a measure of the Occupied Density of States (ODOS), BIS can probe the UDOS. In the case of the actinides there is a further advantage: by working at higher energies, the 6d contributions to the BIS spectrum are suppressed, driven by the relative cross sections of the 5f and 6d states, providing what is often a pure 5f UDOS. [32, 57] This effect can be clearly seen in Figure 1, Panel A, where the U6d peak is observable at 20.7 eV [48], severely diminished at 736 eV [27] and completely gone at 1487 eV. [42]

It is fair question to ask: why probe the unoccupied states if the bonding is actually derived from the filled states? The answer is simple. While the bonding does go through the occupied states, the complementarity of the bonding and anti-bonding or occupied and unoccupied states [58, 59] means that the Unoccupied Density of States (UDOS) provides detailed information about the nature of the 5f electronic structure and thus the bonding. (Figure 1, Panel C) As described briefly above, the problem is that the occupied 5f electronic structure and corresponding photoelectron spectra are all much the same for many different materials. There is not enough spectroscopic structure upon which to differentiate various models. The UDOS provides that detail.

# Determining the j-Specific Unoccupied 5f Electronic Structure in Actinides

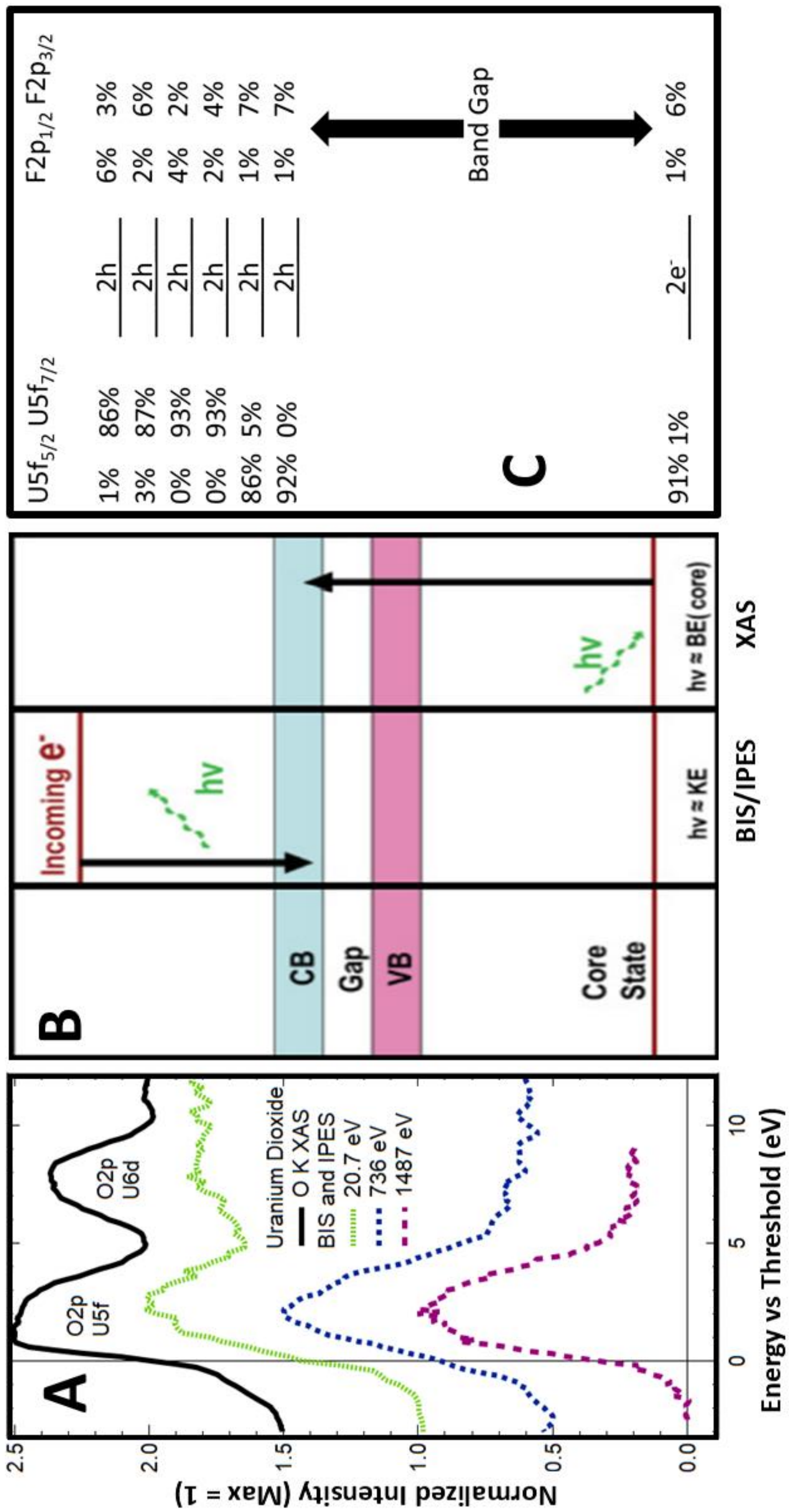

*Figure 1 A: Comparison of Ok XAS and IPES/BIS at different energies for uranium dioxide. The hv = 20.7 eV IPES is from Chauvet and Baptist [48]. The hv = 1487 eV is from Baer et al. [42] B: A schematic of the BIS/IPES and XAS/HERFD processes. For BIS/IPES, hv is the energy of the outgoing photon. KE is the kinetic energy of the incoming electron. For XAS/HERFD, hv is the energy of the incoming photon, which is absorbed. CB is the conduction band. VB is the valence band. [40] C: This is a cartoon illustrating the relationship between occupied and unoccupied states of U 5f origin in uranium tetrafluoride. The percentages are taken from Ryzhkov's calculations [58], using a $UF_8^{4-}$ cluster of $C_{4v}$ symmetry. The larger bandgap is based upon experimental results. [59] Hole = h. Electron = e-. The sums of percentages are less than 100%: there are small mixings from other states, which are neglected here. Uranium tetrafluoride is quite ionic, thus the intermixing with even the O2p states is limited.*

Returning to the issue of 5f selectivity in BIS, the XAS/HERFD (Figure 1) has a similar selectivity: the dominance of electric dipole transitions limits the final states to the 5f and 7p UDOS in the $M_{4,5}$ spectroscopy. [32, 36-40] In the XAS/HERFD spectra below, both 5f and 7p contributions will be seen, but they occur at different energies, thus providing an isolation of the 5f UDOS. Moreover, the electric dipole transitions provide a further selectivity: for the $M_4$ transition, the final state is $5f_{5/2}$ only; for the $M_5$ transition, the final state is mainly the $5f_{7/2}$ manifold. Thus, for the $M_5$ transition, the $5f_{7/2}$ is strongly dominant. [32, 36-40]

As an example of the utility of improved selectivity, consider Figure 2. Here, the spin-resolved photoelectron spectroscopy (SRPES) of $Fe_3O_4$ is shown. [60, 61] Clearly, the two different spins have substantially different Density of States (DOS), as can be seen in Panel A. It is of interest that this spin polarization can be ruined, as illustrated in Panel B. The goal of this work is to demonstrate that HERFD of the $M_{4,5}$ edges in the actinides can function in the same manner as SRPES, proving improved selectivity in the spectroscopic process. However, while the important quantum number in SRPES is the spin. i.e. $m_s = \pm ½$, the important quantum number in the 5f states is the j value, 5/2 or 7/2, corresponding to the final state manifolds of $5f_{5/2}$ and $5f_{7/2}$ natures.

Two examples the correlation between BIS and XAS/HERFD results is presented in Figure 3, using U(IV) [27, 32, 36, 40] and Th(IV) [62,63] oxidized samples. (This nomenclature will be used for oxidation state, with the roman numeral IV indicating a 4+ charge in the fully ionic limit but allowing for the possibility of incomplete oxidation.) Both uranium dioxide [27, 32, 36, 40] and thorium dioxide [62] are fluorite structures, with similar lattice parameters and a cubic environment for the actinide ion surrounded by 8 oxygen nearest neighbors. Interestingly, both are completely ionized with a U electron configuration of U $5f^2(spd)^0$, n = 2 and a Th configuration of Th $5f^0(spd)^0$, n = 0. Here, n is the number of 5f electrons. ($ThF_4$ will also have a completely ionized Th atom. [63]) This means that the U(IV) will have broadening from the final state angular momentum coupling, but the Th(IV) will not. (The excited electron in Th, moved from the 3d to the 5f manifold will have no other electrons with which to couple. It takes two or more electrons for angular momentum coupling.) [32, 36, 40] Note how much narrower the Th(IV) case is than the U(IV) case. These results will be discussed in more detail below. It is of interest that the Th(IV) case is not only narrower than the U(IV) case, it is also narrower than the case for Th metal. In Th metal, the configuration is Th $5f^0(spd)^4$, and angular momentum coupling with the $(spd)^4$ broadens the spectrum there. [33- 35, 43]

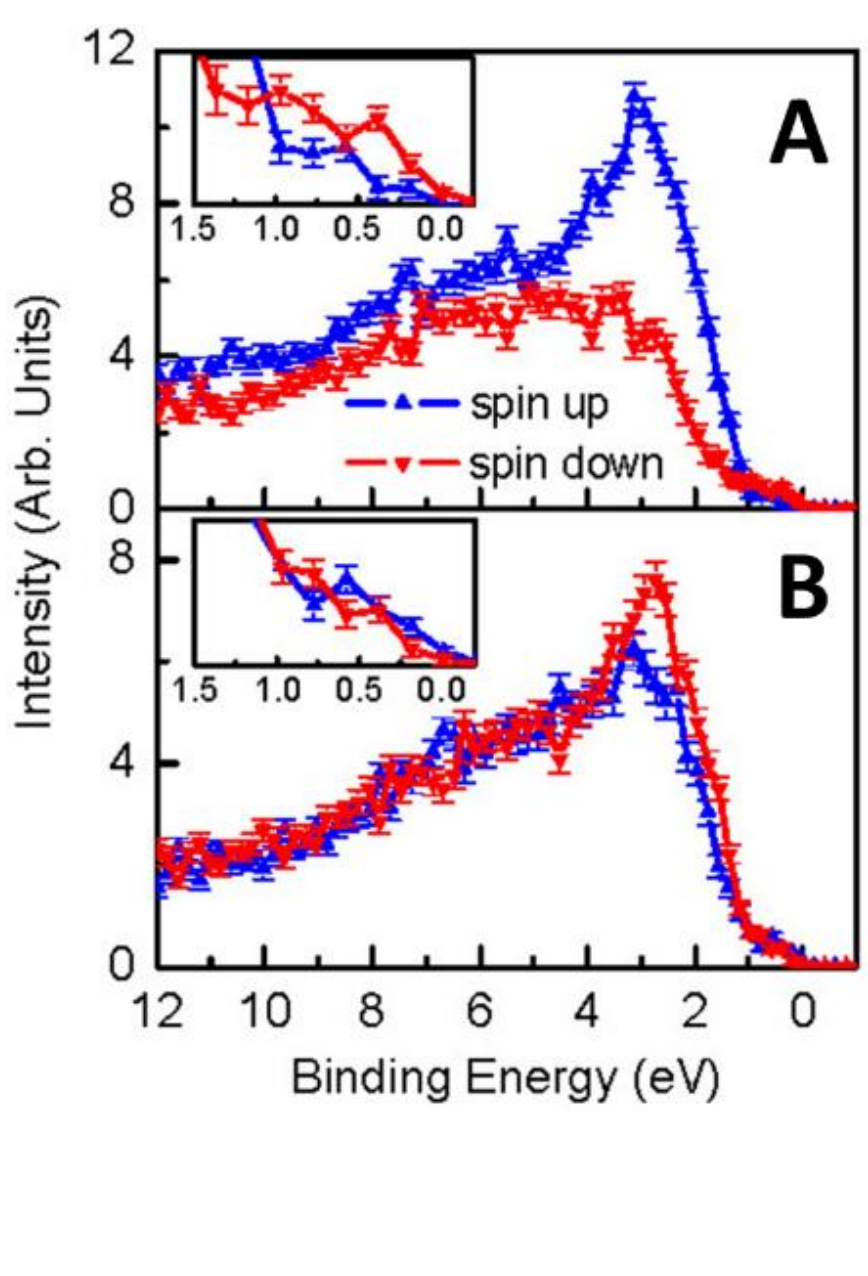

*Figure 2 Spin Resolved PES of $Fe_3O_4$. [60, 61] Panel A: Spin resolved photoemission spectrum from an "as received" sample at hv = 160 eV. Panel B: Spin resolved photoemission spectrum from a damaged sample, i.e. after $Ne^+$ sputtering, at hv = 160 eV. Note the large effect observed in the as received sample in A and the loss of the effect in B.*

# Determining the j-Specific Unoccupied 5f Electronic Structure in Actinides

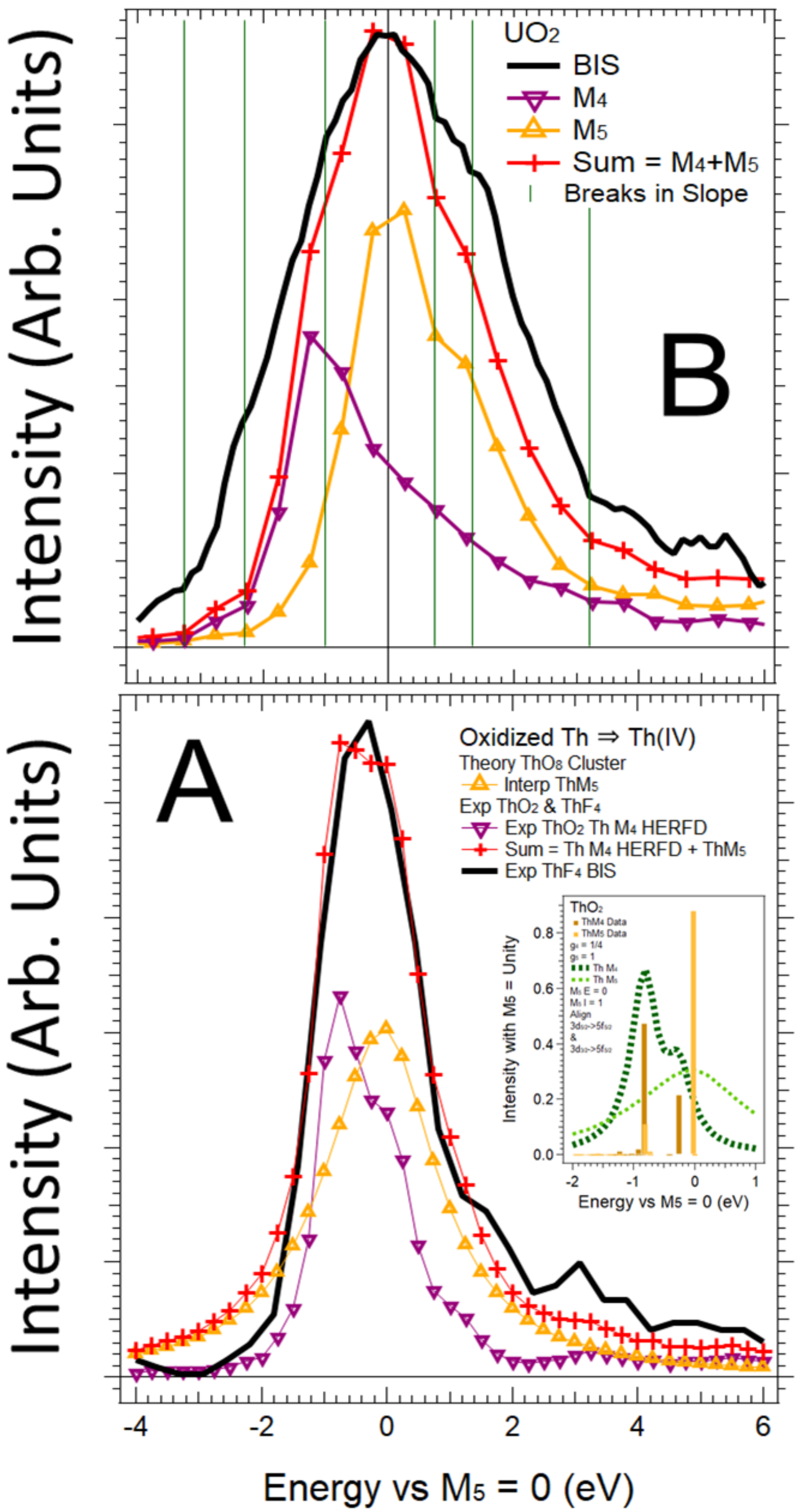

*Figure 3 Top panel B: the results for U(IV), using the XAS/HERFD of $UO_2$ with a 0.47 scaling of $M_4$:$M_5$ EXAFS, from BR = 0.68, and the $UO_2$ BIS spectrum. The vertical green lines denote regions with changes in slope for the HERFD and BIS. The HERFD energy scale was shifted to compare with the BIS data, and HERFD intensity was normalized to the magnitude the BIS scale. The energy of the BIS electron beam was 736eV. The BIS spectrum is taken from Ref. [27]. Bottom Panel A: the results for Th(IV). This comparison uses $ThO_2$ $M_4$ XAS/HERFD [62] and $ThF_4$ BIS [63] experimental results, as well as a theoretically generated $M_5$ XAS/HERFD using a $ThO_8^{12-}$ cluster. [32] The inset shows the alignment in the Th(IV) case using Theory.* [32,40] See the text for further discussion.

# Determining the j-Specific Unoccupied 5f Electronic Structure in Actinides

Here, this new insight that the U 3d HERFD can be used as a direct probe of the $5f_{5/2}$ and $5f_{7/2}$ UDOS will be applied to earlier BIS measurements of the total 5f UDOS. The remainder of the paper is organized as follows. Next, the experimental details will be presented. The Experimental section will lead into a discussion of five main cases. Finally, a summary and conclusion section will be presented at the end.

## II Experimental

The $UO_2$, $UCd_{11}$, and $UPd_3$ HERFD X-Ray Absorption Spectroscopy (XAS) experiments were performed using a Si(111) monochromator and a high-resolution Johansson-type spectrometer [64] operating in the tender x-ray regime (1.5 - 4.5 keV), on Beamline 6-2a at the Stanford Synchrotron Radiation Light-source. (HERFD = High Energy Resolution Fluorescence Detection) The total energy resolution broadening of the HERFD experiment is about 0.8 eV. This produced a sharpening the U $M_4/M_5$ spectroscopic features of almost an order of magnitude when compared to conventional x-ray absorption spectroscopy. Further detail concerning the HERFD experiments is available elsewhere [32, 36, 40, 65].

FEFF has been used in support of the experimental HERFD and BIS measurements. The name FEFF is derived from the term $f_{eff}$ used in x-ray scattering theory. FEFF is a Green's Function spectral simulation program used in x-ray absorption spectroscopy and related techniques, including self-consistent real space multiple-scattering code for simultaneous calculations of x-ray-absorption spectra and electronic structure, particularly for Extended X-ray Absorption Fine Structure (EXAFS). [65- 70]

# Determining the j-Specific Unoccupied 5f Electronic Structure in Actinides

Bremsstrahlung Isochromat Spectroscopy (BIS) is a high energy variant of Inverse Photoelectron Spectroscopy (IPES), with an electron impinging upon a sample and emitting a photon. The final state of the electron is in the manifold of the Unoccupied Density of States. Generally speaking, the photon emission is governed by electric dipole transition selection rules and intensities and the photon energy will be approximately the same magnitude as the initial kinetic energy of the incoming electron. (Figure 1) More specifically, in metallic systems, KE = hv - $H^F$ + ϕ, where KE is the kinetic energy, hv is the energy of the outgoing photon, H is the energy of the hole relative to the Fermi Energy and ϕ is the work-function. [70] Usually, $\phi_{effective}$ - $H_{effective}$ is small, such that hv ≈ KE. Most BIS and IPES measurements are done at a fixed hv and the KE is varied. The LLNL BIS measurements were performed using multichannel detection, fixing KE and taking a snapshot across a finite energy range, with many energy channels collected simultaneously. [70,71] Thus, the $UO_2$ BIS experiment at KE = 736 eV was carried out in house at LLNL using a multichannel, monchromatized photon detection scheme. [27, 71-73] The Baer BIS experiments for UAs, $UBe_{13}$ and $UPd_3$ were carried out at hv = 1487 eV ($AlK_\alpha$ energies) and with a bandpass of about 1 eV. More detail concerning these measurements can be found in the source articles. [41-44] BIS has an excellent potential to probe U5f states, in part because of the dominance at these energies of the 5f cross sections over those of other possibilities, such as the U6d. Foundational information about cross sections can be found elsewhere [57], but must be corrected for the impact of initial and final state occupations/unoccupations in each specific spectroscopy. [32, 37-39] Often in BIS, the strong, sharp spectral structure will come from f and d manifolds, with the more diffuse and weaker s and p contributions being neglected.

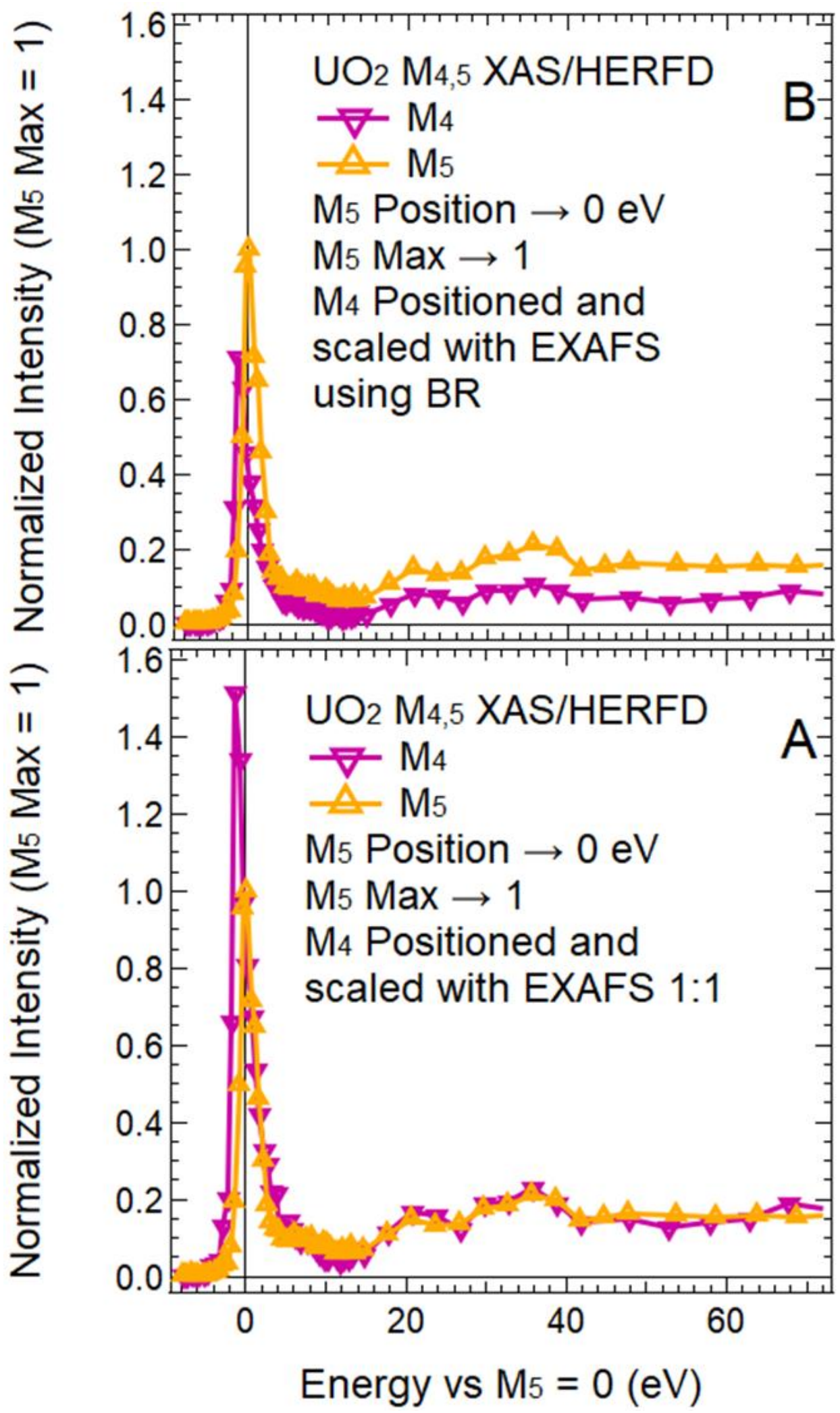

*Figure 4 Shown here is the experimental alignment and scaling of the XAS/HERFD pairs, using EXAFS and the Branching Ratio (BR). The $M_5$ spectrum is normalized in height and set to zero on the energy axis. (A) $M_4$ with EXAFS and 1:1 vertical scaling. (B) $M_4$ with EXAFS and vertical scaling of 0.47 from the BR = 0.68. BR = $I_5/(I_4 + I_5)$. For n =2 and UO2, it has been demonstrated that BR = 0.68. [24-26, 32] If BR = 0.68, $I_4/I_5$ = 0.47. Note that the separation of maxima is about 1 eV, consistent with previous determinations of the spin-orbit splitting [11, 25, 32, 36]*

Before proceeding, it is useful to address the issue of the spin-orbit splitting and scaling in the $M_{4,5}$ XAS/HERFD spectra. Consider Figure 4, where the $M_4$ and $M_5$ XAS/HERFD spectra of $UO_2$ are shown, aligned and scaled on the EXAFS features near 20 and 40 eV. This procedure produces a spin-orbit splitting of about 1 eV, consistent with earlier determinations. [11, 25, 32, 36] This scaling procedure is based upon the past observation that the plateau heights that follow asymmetric peaks are proportional to the peak areas. [74, 75]

## III Results and Discussion

Now, let us turn our attention to a discussion of the relevant examples.

### IIIa Actinide Dioxides: $ThO_2$ and $UO_2$

Consider again the results for $UO_2$ and $ThO_2$ shown in Figures 3 and 4. In the X-Ray Absorption process, a photon is annihilated and an electron moves from the 3d core level to the 5f UDOS; $3d^{10} + 5f^{n} (spd)^{0} + h\nu \rightarrow 3d^{9} + 5f^{n+1}(spd)^{0}$, for $UO_2$ with n = 2 and $ThO_2$ and $ThF_4$ with n = 0. In the final state, it takes at least two electrons to have angular momentum coupling. (Of course, filling of the 5f levels completely or either of the $5f_{5/2}$ or $5f_{7/2}$ substructures can also quench angular momentum coupling. [4, 23, 25, 28, 76]) If n ≥ 1, there should be angular momentum coupling in the final state and this is observed both experimentally and theoretically for $UO_2$ (n = 2) and $PuO_2$ (n = 4). [32, 40] However, for Th (IV) and complete ionization, n = 0 and n + 1 = 1, so there will be no angular moment coupling broadening in the $ThO_2$ and $ThF_4$. The breaks in slope in the $UO_2$ BIS spectrum in Figure 3 are the manifestation of the underling structure driven by angular momentum coupling, agreeing with the predictions of computational theory. [32] There is a high level of agreement here between the XAS/HERFD and BIS results for $UO_2$, with the strong indication that the j-substructure is actually being resolved.

# Determining the j-Specific Unoccupied 5f Electronic Structure in Actinides

In the case of the $UO_2$, the alignment and scaling of the $M_4$ and $M_5$ spectra is done completely based upon experiment, using the EXAFS peaks and the known Branching Ratio. This is presented in Figure 4 and discussed in the Experimental Section above. For the Th (IV) case, the data set was less complete and the alignment is made based upon the theoretical modelling [32], as illustrated in the inset of Panel A in Figure 3. The scaling was achieved by applying the BR = 0.6 ($I_4/I_5$ = 2/3, with Ix being the integrated intensity of the x peak) to the $M_4$ experimental data and the $M_5$ theoretical prediction. For it is known that for Th and n = 0, BR = 0.6 [24-26, 32] Despite using an admixture of $ThO_2$ data ($M_4$), $ThO_8^{12-}$ cluster theory ($M_5$) and $ThF_4$ data (BIS), there is again remarkable agreement between the XAS/HERFD and the strong indication that the $5f_{5/2}$ and $5f_{7/2}$ substructure is being resolved.

The actinide dioxides have proven to be a potent argument in support of the central hypothesis of this work. However, as can be seen in Figure 2, if a strong effect can be achieved, should it not also be possible to destroy the effect? This question will be addressed next.

**IIIb Uranium Monopnictides from Semimetals: UAs and USb**

It is well known that the uranium mono-chalcogenides (US, USe, UTe) and mono-pnictides (UN, UP, UAs, USb) exhibit long range magnetic ordering, with the monochalcogenides being ferromagnetic and the mono-pnictides being antiferromagnetic. [77]. In the limit of complete ionization, one would expect the uranium ion to have a configuration of $U^{2+}5f^3(7s7p6d)^1$ for the monochalcogenides and $U^{3+}5f^3$ for the mono-pnictides. For UAs and USb, the ligand d states should not be a factor in the UDOS: the valence d states are present as completely filled shallow cores. This assertion is confirmed by the FEFF calculation displayed in Figure 5. Similar calculations for US and UTe can be found in an earlier publication. [33] Thus for these cases, the BIS will be completely dominated by the U5f contributions.

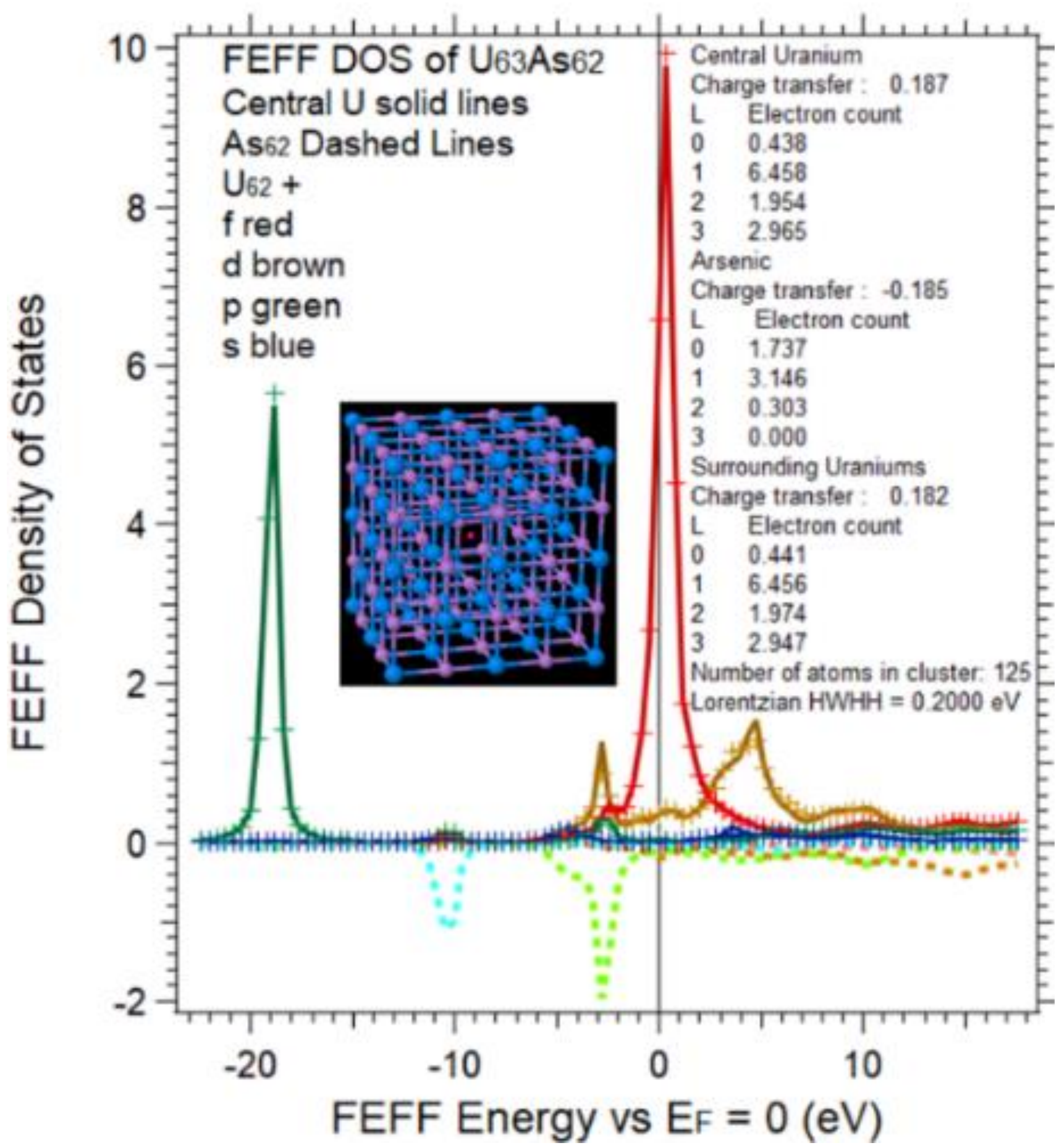

*Figure 5 Here is shown the results of FEFF DOS calculation for a $U_{63}As_{62}$ cluster, in a NaCl rock-salt structure, as shown in the inset. The central U atom is represented by a red sphere; the As atoms as pink spheres and the surrounding U atoms as blue spheres.*

One way to quench the spin-orbit polarization is to cause the states to mix. In a strongly magnetic system with a high degree of order, the states will be dominated by the magnetic effects and perturbed by the crystal field effects, with the $5f_{5/2}$ and $5f_{7/2}$ potentially mixing thoroughly, if not completely. [78, 79] Under these conditions, one might expect the $M_4$ XAS/HERFD and $M_5$ XAS/HERFD and BIS to look much the same. That is what is observed in Figure 6. [46, 80] In each, there is a central maximum with a higher and lower energy shoulder. The lower energy shoulder in the BIS is clipped by the Fermi Level, making it look more abrupt. The higher energy shoulder in the BIS is probably also riding on the inelastic scattering endemic to electrons, with possibly additional contributions of the U 7s, U7p and U6d and As states, as illustrated in Figure 5. Nevertheless, all are much the same, suggesting strong $5f_{5/2}$ and $5f_{7/2}$ mixing.

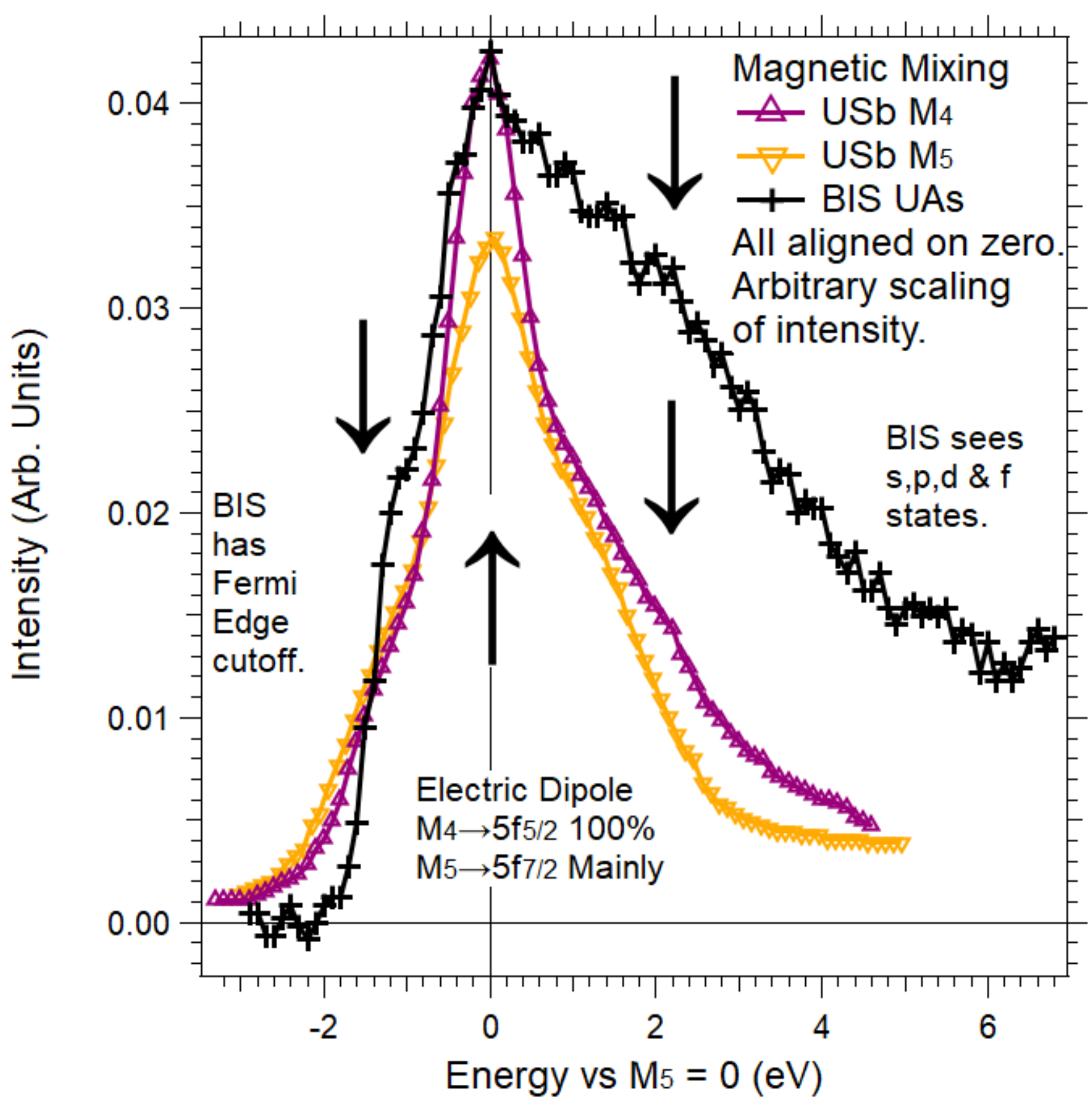

Figure 6 Presented here is a comparison of the $M_{4,5}$ XAS/HERFD of USb (K. Kvashnina et al, with permission [80]) and the BIS of UAs. [14, 46] Here, all three have their maximum aligned at zero and are arbitrarily scaled vertically.

# Determining the j-Specific Unoccupied 5f Electronic Structure in Actinides

Now, consider the case of isolated U atoms.

**IIIc Simple 5f localized case with n = 3: $UCd_{11}$, and $UBe_{11}$**

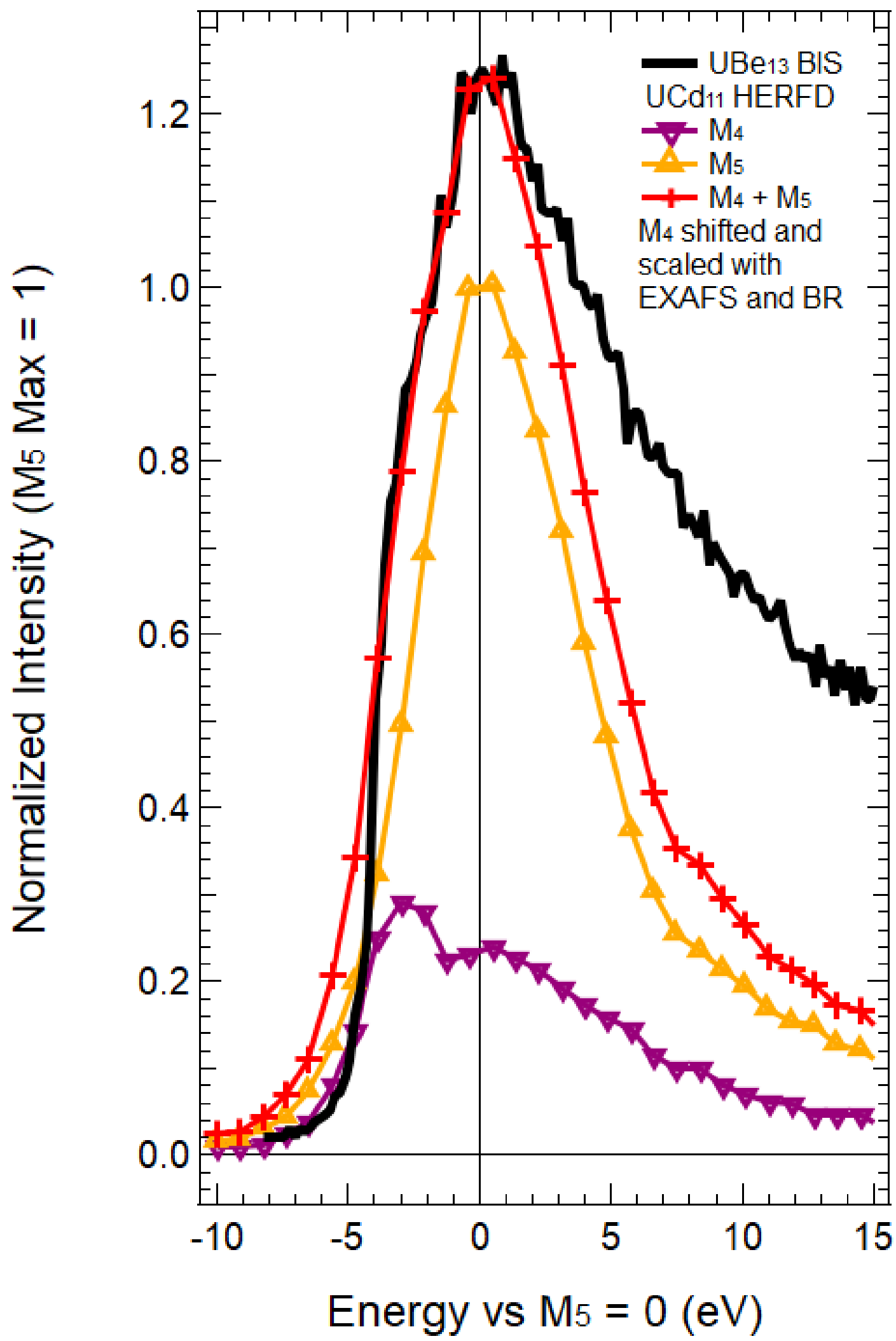

*Figure 7 Shown here is a comparison of the $UBe_{13}$ BIS and the $UCd_{11}$ HERFD results. For the HERFD scaling, the BR value of 0.72 has been used, producing an $I_4/I_5$ ratio of 0.39. See the text for further detail. The BIS spectrum is from Baer et al. [45]. In the Intermediate Coupling Model, for a localized, n = 3 case, BR = 0.72. [24-26, 32]*

A previous analysis of the HERFD of $UF_4$ and $UCd_{11}$ indicated that the U in $UF_4$ was a localized n =2 and the U in $UCd_{11}$ was a localized n = 3, consistent with the Intermediate Coupling Model (ICM). [36] This has been confirmed by further experimental work. [81] The ICM is an atomic model without further complications or 5f mixing. It has been shown [82] that

# Determining the j-Specific Unoccupied 5f Electronic Structure in Actinides

$UCd_{11}$ consists of widely spaced U atoms ($r_{UU}$ = 6.56Å), surrounded by Cd atoms, with 12 nearest-neighbors ($r_{UCd}$ = 3.45Å) and 8 next-nearest neighbor Cd atoms, in a cubic $BaHg_{11}$ structure (space group: Pm-3m). This type of caged structure is consistent with a picture of an isolated U atom with a localized but not strongly ionized $5f^3$ configuration. Again, more recent results [81] indicate that $UCd_{11}$ is $5f^3$ and localized.

$UBe_{13}$ is a good candidate for comparison to $UC_{11}$. The structure of the unconventional superconductor $UBe_{13}$ is not the same as $UCd_{11}$, but with strong parallels. Amon et al. [83] report that the $UBe_{13}$ is a cubic $NaZn_{13}$ structure type with space group Fm3̌c and a lattice parameter of 10.254 Å. The Uranium at the 8a site is caged by 24 Be atoms. It is also important that the electronic configuration of Be is $1s^2 2s^2$, with only small cross section 2s and 2p empty states and no potential for significant Be d or Be f contributions to the UDOS. (Even with corrections for stoichiometry and the number of unoccupied vs occupied states, the Be 2s and 2p cross sections are still small.) Thus, it is expected that the BIS of $UBe_{13}$ will be dominated by the U5f UDOS due to the sharpness and large magnitude of the 5f cross sections, relative to other potential contributions from either U or Be. This expectation is confirmed in Figure 7.

Aside from the stronger tailing (E > 0) in the BIS, there is excellent agreement between the XAS/HERFD and the BIS. Again, it is reasonable to expect that the electrons in BIS will generate more secondary electrons than the photons in HERFD, i.e. the peak asymmetry will be stronger in BIS than HERFD. (This is consistent with the much longer mean free paths for photons than electrons of comparable energy. [65, 84] However, it should be noted that the Fermi Energy cutoff in the BIS is not reproduced in the HERFD analysis. Of course, it is not expected that the HERFD would exhibit a Fermi Edge, owing to the residual lifetime broadening,

There is a nagging question whether it is appropriate to think of a correlated electron and unconventional superconducting material such as $UBe_{13}$ in such a simple way. [85] However, there is an interesting counter example. The high temperature superconductor $YBa_2Cu_3O_{6.9}$ has been shown to possess an electronic structure that agreed in great detail with conventional calculations. [86] This result suggests that the collective effects are layered on top of what would be a fairly normal electronic structure, in support of validity of the comparisons made above, particularly since the BIS measurements were not made at low temperatures.

Having succeeded with the simple case U $5f^3$ localized systems, a more complicated example is considered next, where the ligand d states can strongly affect the BIS results.

**IIId UPd3: U5f and Pd4d**

Now, consider a slightly more complicated case, $UPd_3$ as illustrated in Figure 8. Here, it is clear that the U $M_{4,5}$ HERFD is not sufficient to explain the BIS. Instead, there is strong evidence that the Pd 4d states are present in the BIS as well.

In the case of the U $M_{4,5}$ of $UPd_3$, a BR of 0.6 was used, which gives a scaling of $I_4/I_5$ = 2/3. A BR = 0.6 implies a statistical filling of the 5f states, but with the retention of a spin-orbit splitting between the $5f_{5/2}$ and $5f_{7/2}$ manifolds. It generates a relatively slightly larger $M_4$ peak than the larger BR values used earlier. The positioning and scaling of the $M_{4,5}$ spectra has been done using the EXAFS at higher energies and the spin-orbit splitting remains about 1 eV, which is in good agreement with the calculations of Yaresko et al., as shown in Figure 8. [87] The $M_4$ and $M_5$ spectra are more dissimilar than was observed in the USb case, so the mixing is less severe.

# Determining the j-Specific Unoccupied 5f Electronic Structure in Actinides

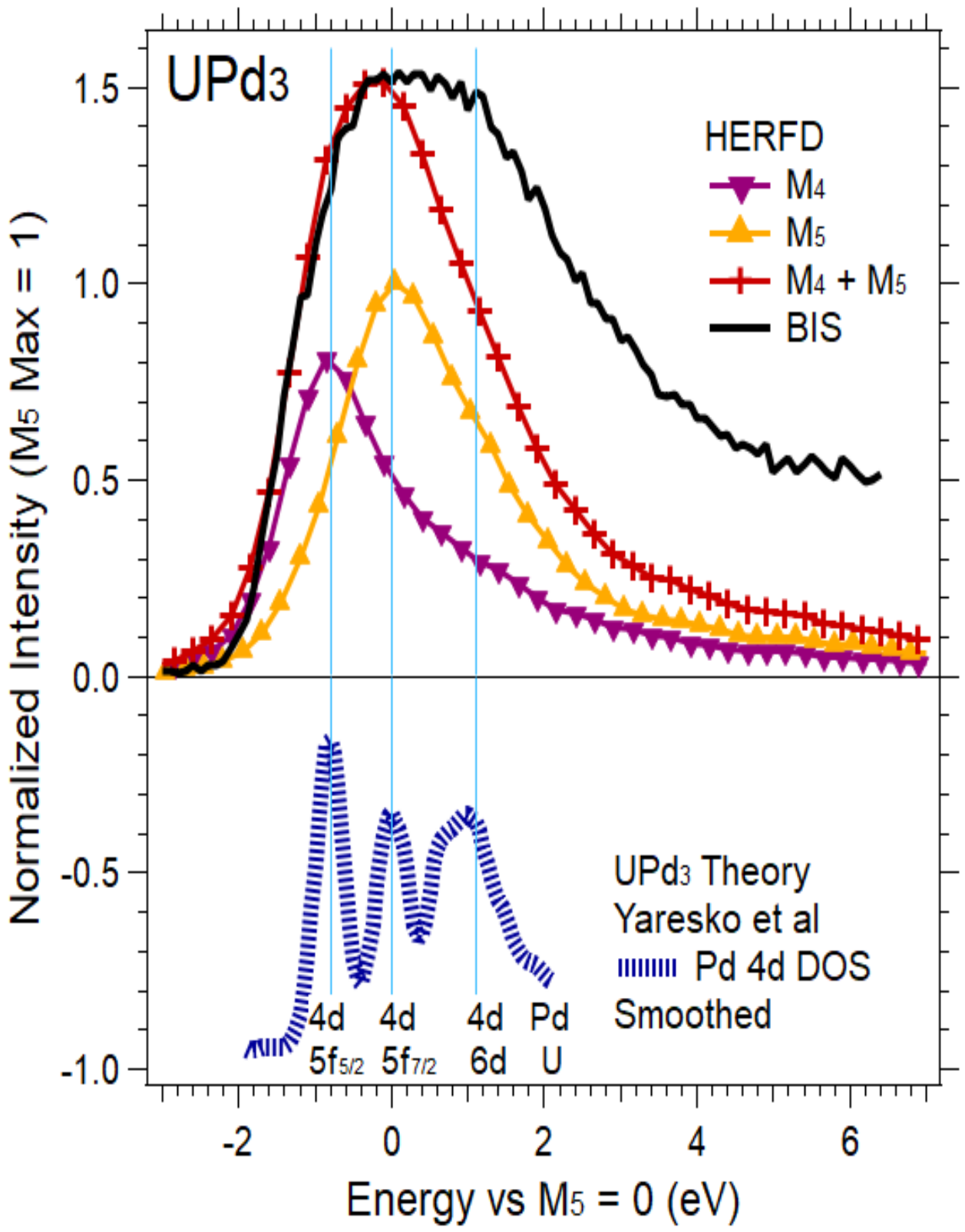

*Figure 8 Shown here is a comparison of the HERFD, BIS and Pd 4d DOS for $UPd_3$. The Pd 4d DOS theory from Yaresko [87], with permission. The BIS is from Baer [14, 44, 46] For the HERFD, three curves are presented: the $M_4$ HERFD spectrum (purple), the $M_5$ HERFD spectrum (orange) and the sum of the $M_4$ and $M_5$ (red). The black line represents the BIS data. The $M_5$ spectrum has been normalized (Max = 1) and placed at zero eV. The $M_4$ spectrum has been shifted and scaled using the EXAFS and BR method, using BR = 0.6 and $I_4/I_5$ = 2/3, the statistical limit. The BIS is arbitrarily shifted and scaled, to enhance the match near the edge jump and at the highest energies. The Pd 4d theory (blue slashes) has been shifted and scaled arbitrarily, but as with the BIS, the energy steps remain intact. The Pd 4d theory has been smoothed, undergoing a 5 point smooth twice.*

The theory from Yaresko et al. [87] exhibits three lobes in the Pd UDOS which line up with $U5f_{5/2}$, $U5f_{7/2}$ and U6d UDOS. Roughly similar computational results for $UPd_3$ have also been reported by Eriksson et al. [88] BIS is not sensitive to the U6d but would pick up the Pd4d. The U6d-Pd4d interaction appears to be stronger than U5f-Pd4d interaction, perhaps suggestive of a directional covalent bond. [87] Yaresko et al. also report that there are two U sites in unit cell. [87] This could also be a source of broadening in the results. While it is clear that the Pd 4d states play an important role in the BIS of $UPd_3$, the $M_{4,5}$ HERFD does correctly project the j specific 5f contributions from U.

## IV Summary and Conclusions

An empirical comparison of high resolution $M_{4,5}$ HERFD/XAS with Bremsstrahlung Isochromat Spectroscopy (BIS) of various actinide materials has been presented here. It has been demonstrated that XAS-HERFD is essentially the j-specific version of BIS, providing a direct measure of the separate U $5f_{5/2}$ and U $5f_{7/2}$ Unoccupied Density of States (UDOS).

**Acknowledgements**

Stanford Synchrotron Radiation Light-Source is a national user facility operated by Stanford University on behalf of the DOE and the OBES. Partial funding for the instrument used for this study came from the U.S. Department of Energy, Office of Energy Efficiency & Renewable Energy, Solar Energy Technology Office BRIDGE Program. Resources of the National Energy Research Scientific Computing Center, a DOE Office of Science User Facility supported by the Office of Science of the U.S. Department of Energy under Contract No. DE-AC02-05CH11231, were used in this work. J.G.T. wishes to thank S. Nowak, C.H. Booth, E.D. Bauer, S.-W. Yu, R. Alonso-Mori, T. Kroll, D. Nordlund, T.-C. Weng and especially D. Sokaras. J.G.T. also wishes to thank the University of Wisconsin-Oshkosh and its faculty and staff for their ongoing support and encouragement. It is a gem set beside the Fox River and Lake Winnebago.

**The author has no conflict of interests.**

**Data is available upon reasonable request.**

# Determining the j-Specific Unoccupied 5f Electronic Structure in Actinides